 \documentclass[12pt,preprint]{aastex}

\shorttitle{Study of EUV emission and properties of a coronal streamer}
\shortauthors{Goryaev et al.}

\begin{document}


\title{Study of EUV emission and properties of a coronal streamer from PROBA2/SWAP, Hinode/EIS and Mauna Loa Mk4 observations}


\author{F.~Goryaev\altaffilmark{1}, V.~Slemzin, L.~Vainshtein}
\affil{P.N. Lebedev Physical Institute of the RAS (LPI), Moscow, Russia 119991}

\and

\author{David R.~Williams}
\affil{Mullard Space Science Laboratory, University College London, Holmbury St Mary, Surrey, RH5 6NT, U.K.}




\altaffiltext{1}{e-mail address: goryaev\_farid@mail.ru}



\begin{abstract}
Wide-field EUV telescopes imaging in spectral bands sensitive to 1~MK plasma on the Sun often observe extended ray-like coronal structures stretching radially from active regions to distances of 1.5--2$R_{\odot}$, which represent the EUV counterparts of white-light streamers. To explain this phenomenon, we investigated the properties of a streamer observed on October 20--21, 2010 by the {\sl PROBA2}/SWAP EUV telescope together with the {\sl Hinode}/EIS spectrometer (HOP 165) and the Mauna Loa Mk4 white-light coronagraph. In the SWAP 174~\AA\ band comprising the Fe~{\sc ix} -- Fe~{\sc xi} lines, the streamer was detected to a distance of 2$R_{\odot}$. We assume that the EUV emission is dominated by collisional excitation and resonant scattering of monochromatic radiation coming from the underlying corona. Below 1.2$R_{\odot}$, the plasma density and temperature were derived from the {\sl Hinode}/EIS data by a line-ratio method. Plasma conditions in the streamer and in the background corona above 1.2$R_{\odot}$ from disk center were determined by forward-modeling the emission that best fit the observational data in both EUV and white light. It was found that plasma in the streamer above 1.2$R_{\odot}$ is nearly isothermal, with a temperature $T=1.43\pm 0.08$~MK. The hydrostatic scale-height temperature determined from the evaluated density distribution was significantly higher (1.72$\pm$0.08~MK), which suggests the existence of outward plasma flow along the streamer. We conclude that, inside the streamer, collisional excitation provided more than 90\% of the observed EUV emission; whereas, in the background corona, the contribution of resonance scattering became comparable with that of collisions at $R \gtrsim 2R_{\odot}$.
\end{abstract}




\section{Introduction}

Extreme Ultraviolet (EUV) imaging of the inner solar corona gives the most valuable information about the spatial and temporal dynamics of coronal plasma, because this spectral range corresponds to its inherent temperature range of 1--2~MK. The temperature sensitivity of the EUV emission results in noticeable differences in coronal structures that are seen simultaneously in the EUV and white-light spectral ranges during eclipses \citep{Pasachoff11}. Whereas in white light the corona is routinely studied at distances from disk center (heliocentric distance) exceeding 2$R_{\odot}$, in EUV the corona is typically observed below 1.3$R_{\odot}$, because ordinary EUV telescopes and spectrometers have a limited field of view adapted for imaging of the solar disk with the highest possible resolution. Examples of such studies can be found elsewhere \citep[see, e.g.,][and others]{Gibson99,Feldman99,Parenti00,Warren02,Curtain06,Schmit11,Kucera12}. At distances of 1.3--11$R_{\odot}$, systematic spectroscopic investigations of the coronal plasma in the selected EUV lines were performed with the Ultraviolet Coronagraph Spectrometer \citep[UVCS;][]{Kohl95} on board the {\sl Solar and Heliospheric Obseratory} ({\sl SOHO}). Although this instrument observed the corona only in a scanning mode, it gave significant results in the study of properties of various coronal structures, such as active region streamers, quiescent equatorial streamers, polar coronal plumes, {\it etc.} \citep[and references therein]{Kohl06}. In particular, it was found that plasma in solar-minimum streamers is in thermal equilibrium, with the electron, proton and ionic temperatures all at similar values between 1.1 and 1.5~MK.

The first real imaging of the extended EUV corona above 1.3$R_{\odot}$ at high solar activity was performed in 2002 with the SPectrographIc X-Ray Imaging Telescope-spectroheliograph (SPIRIT) onboard the {\sl Complex ORbital Observations Near-Earth of Activity of the Sun} ({\sl CORONAS-F}) satellite in its coronagraphic mode \citep{Slemzin08}. The telescope operated in the 175~\AA\ spectral band, which comprises the resonance lines of Fe~{\sc ix} to Fe~{\sc xi} with excitation temperatures around 1~MK. In several dedicated coronagraphic observations, it was found that the EUV corona contained elongated bright ray-like structures stretching from some active regions to distances of 2--3$R_{\odot}$, with brightness 2--4 times greater than that of the ambient background corona. Comparison has shown that coronal structures seen by SPIRIT were different from those observed in UVCS scans observing the colder O~{\sc vi} line, probably due to differences in the emission mechanisms or excitation temperatures. Similar structures have been detected during eclipses in visible lines generated in transitions between higher levels of the same Fe~{\sc ix} -- Fe~{\sc x} ions \citep{Habbal11}. Recently, it was shown that these ray-like structures very likely represent coronal signatures of long living plasma outflows originating at the interfaces between active regions and mid-latitude coronal holes, and contributing to the quasi-stationary component of the slow solar wind \citep{Slemzin12}. \citet{Baker09} and \citet{DelZanna11} have shown that outflows are associated with interchange reconnection between closed field lines of active regions and the surrounding open field lines, which introduces a pressure imbalance that drives the plasma outward. Based on this conclusion, we conjecture that plasma outflows propagating along open field lines produce enhancements of density in the corona, which can be seen at the limb as bright, elongated, ray-like structures. In the work of \citet{Slemzin12}, and in the present work, it is shown that these structures correspond spatially to streamers observed above the same active regions by white-light coronagraphs. Therefore, the ray-like structures concerned can be regarded as the EUV counterparts of white-light streamers, revealing inner structure that is indistinguishable in visible light.

Since its launch on November 2, 2009, regular EUV observations of the corona have come from the Sun Watcher with Active Pixels and Image Processing (SWAP) telescope aboard the {\sl Project for Onboard Autonomy 2} ({\sl PROBA2}) mission \citep{Seaton12}. This telescope has a single spectral channel at 174~\AA , with high sensitivity to EUV emission of the same Fe lines as SPIRIT. The extended corona is imaged by SWAP in a special off-pointing mode by co-adding several dozen consecutive frames that are obtained with a cadence of 20--30 seconds. The SWAP telescope has important advantages in studying the corona over similar EUV instruments -- {\it e.g.}, {\sl SOHO}'s EUV Imaging Telescope \citep[EIT;][]{Delab95} and the {\sl Solar Dynamics Observatory}'s Advanced Imaging Assembly \citep[{\sl SDO}/AIA;][]{Lemen12} -- because it has a wide field of view and low stray light, which allow one to see the corona up to 2$R_{\odot}$ and beyond. Regular imaging of the Sun in this off-pointing mode has shown that, even at low solar activity, the corona often displays large-scale EUV structures above active regions with some rays extending up to 2$R_{\odot}$. A photometric analysis of the SWAP data assists in understanding plasma conditions and the physical mechanisms of coronal EUV emission, opening up the possibility of more accurately modeling these.

The role of various mechanisms of EUV emission in the extended corona is still not well established due to a lack of observational data on plasma densities and temperatures at these larger distances, which are needed for numerical modeling of coronal EUV flux. Commonly, it is supposed that the EUV emission of the corona is formed only by collisional excitation \citep[see, e.g.,][]{Golub10}, which results, in particular, in widespread use of the emission-measure approach. By default, the CHIANTI package \citep{Dere97} calculates intensities of the aforementioned Fe ion lines in the corona by taking into account only the collisional excitation mechanism.

Radiative excitation, namely photoexcitation by continuous radiation from the solar disk, is commonly considered only for visible and infra-red lines \citep{Habbal07}. Based on analysis of the 2010 eclipse observations, \citet{Habbal11} determined that radiative excitation by photospheric continuum emission from the solar disk dominates the generation of the monochromatic radiation in the visible Fe-ion lines at heliocentric distances as large as 3$R_{\odot}$. These authors also compared their results with observations in the Proba2/SWAP 174~\AA\ EUV channel, and remarked that collisional excitation is the only excitation process for the Fe~{\sc x} 174~\AA\ emission since the solar disk has practically no emission shortward of 1000~\AA\ corresponding to the wavelength range of these Fe ion lines.

However, the Fe~{\sc ix} -- Fe~{\sc xi} ions in the upper corona can absorb and re-emit the strong monochromatic EUV radiation of those same ions, generated lower in the corona. This emission corresponds to resonance transitions of the ions, which makes excitation and emission by resonant scattering considerably more effective than excitation by continuum radiation from much colder photosphere \citep[see, e.g.,][]{Gabriel71,Withbroe82,Noci87}. It should be noted that this efficiency depends to some degree on the matching of spectral profiles between the illuminating and scattered radiation: in particular, this matching decreases in the case of radially-moving coronal plasma, which results in the Doppler dimming effect.

The collisional excitation rate of a spectral line is proportional to the product of the ion and electron densities, $n_z n_e$, whereas the resonant radiative excitation rate is proportional to only the ion density, $n_z$. This means that at large distances from the solar surface, where plasma density decreases by 2--3 orders of magnitude, the contribution of the resonant scattering in the EUV emission may become comparable or even dominant. Even near the limb, where the electron density is relatively high, the resonant scattering contribution to the radiance of the corona along the line of sight can produce noticeable opacity effects, such as the distortion of intensities and profiles of strong EUV and X-ray spectral lines that are often used in spectroscopic diagnostics (see, e.g., earlier works of \citet{Acton78}, \citet{Schrijver00}, and more recent works of \citet{Antonucci05}, \citet{Mierla08}, \citet{Andretta12}). Relevant estimates of contributions of collisional and radiative excitation mechanisms in emission of the extended corona depend on the radial distribution of electron density, temperature and bulk velocity in coronal structures, which may be established by a detailed spectroscopic consideration of the observational data.

By contrast, white light coronal emission is caused by the Thomson scattering of continuum radiation from the photosphere by free electrons in the corona. The intensity of this emission is proportional to the first power of density, so this emission decreases with distance more slower than the EUV emission and does not depend on plasma temperature.

This paper presents the results of a photometric study of a streamer observed {\bf in} October 2010 with the SWAP instrument in the 174~\AA\ spectral band. On October 20--21 the streamer was seen at the western limb above active region AR 11112, stretching to distances of more than 2$R_{\odot}$. On October 21, the base of the streamer -- from the limb to a distance of 1.2$R_{\odot}$ -- was observed by the EUV Imaging Spectrometer (EIS) on board the {\sl Hinode} spacecraft \citep{Culhane07} in a coordinated EIS/SWAP {\sl Hinode} observing programme (known as HOP 165). On October 20 the same streamer was observed by the Mauna Loa Mk4 coronagraph, which measured the polarization brightness, $pB$, in the streamer produced by the Thomson scattering of photospheric continuum radiation by free coronal electrons.

Based on the analysis of probable excitation mechanisms, we built a forward model of the coronal emission using assumed radial distributions of electron density and temperature in the streamer and in the surrounding corona. The parameters of these distributions were then determined in a self-consistent manner, using a solution that gave the best fit of the simulated EUV emission to the measured SWAP and EIS data, as well as the best fit of the simulated white-light brightness to the $pB$ data provided by the Mk4 coronagraph. Finally, we estimated the relative contribution of resonant scattering in the formation of coronal EUV emission for the plasma conditions determined along the streamer.

Observations of the corona with SWAP are described in Section 2. In Section 3, we consider mechanisms of formation of the coronal EUV emission in Fe ion spectral lines that appear in the SWAP wavelength band. In Section 4, we describe physical conditions in the inner corona and basic relations for modeling the coronal EUV emission. Section 5 describes a determination of the model fitting parameters and comparison of the calculated emission with measurements. Section 6 contains a summary and our conclusions.


\section{Observations and data reduction}\label{sect2}

The SWAP EUV telescope was launched on-board the ESA {\sl PROBA2} microsatellite on 2 November 2009. It provides images of the solar disk and corona in a single spectral band centered on 174~\AA\ over a 54$\times$54 arcmin$^2$ field of view with 3.17-arcsec pixels. The response function of SWAP covers the four most intense spectral lines of Fe~{\sc ix}\,--\,Fe~{\sc xi} ions: Fe~{\sc ix} $\lambda$171.08, Fe~{\sc x} $\lambda$174.53, Fe~{\sc x} $\lambda$177.24, Fe~{\sc xi} $\lambda$180.41. To suppress stray light, the optical two-mirror design includes special baffles. The details of the design and parameters of SWAP are described by \citet{Seaton12}.

SWAP observes the extended corona in a special ``paving'' mode when the satellite offsets the telescope main axis in (heliocentric) X, Y or in both directions, at a fixed angle of 10 arcmin from disk center, to offset the field of view toward the desired part of the corona. The whole corona is imaged as a mosaic sequentially by four offsets: in northeast, southeast, northwest and southwest directions. Images are typically acquired during two 99-min orbits. In order to prevent the undesirable illumination of the pointing system's star trackers by the Earth, the satellite performs four Large Angle Rotations (LARs) during its orbit, each of 90$^\circ$ around the axis directed to the Sun. At each LAR position, observations last about 20~min, with a cadence of 20--30~s thus giving a total of 80--90 images in one orbit.

\begin{figure}    
\epsscale{0.9}
\plotone{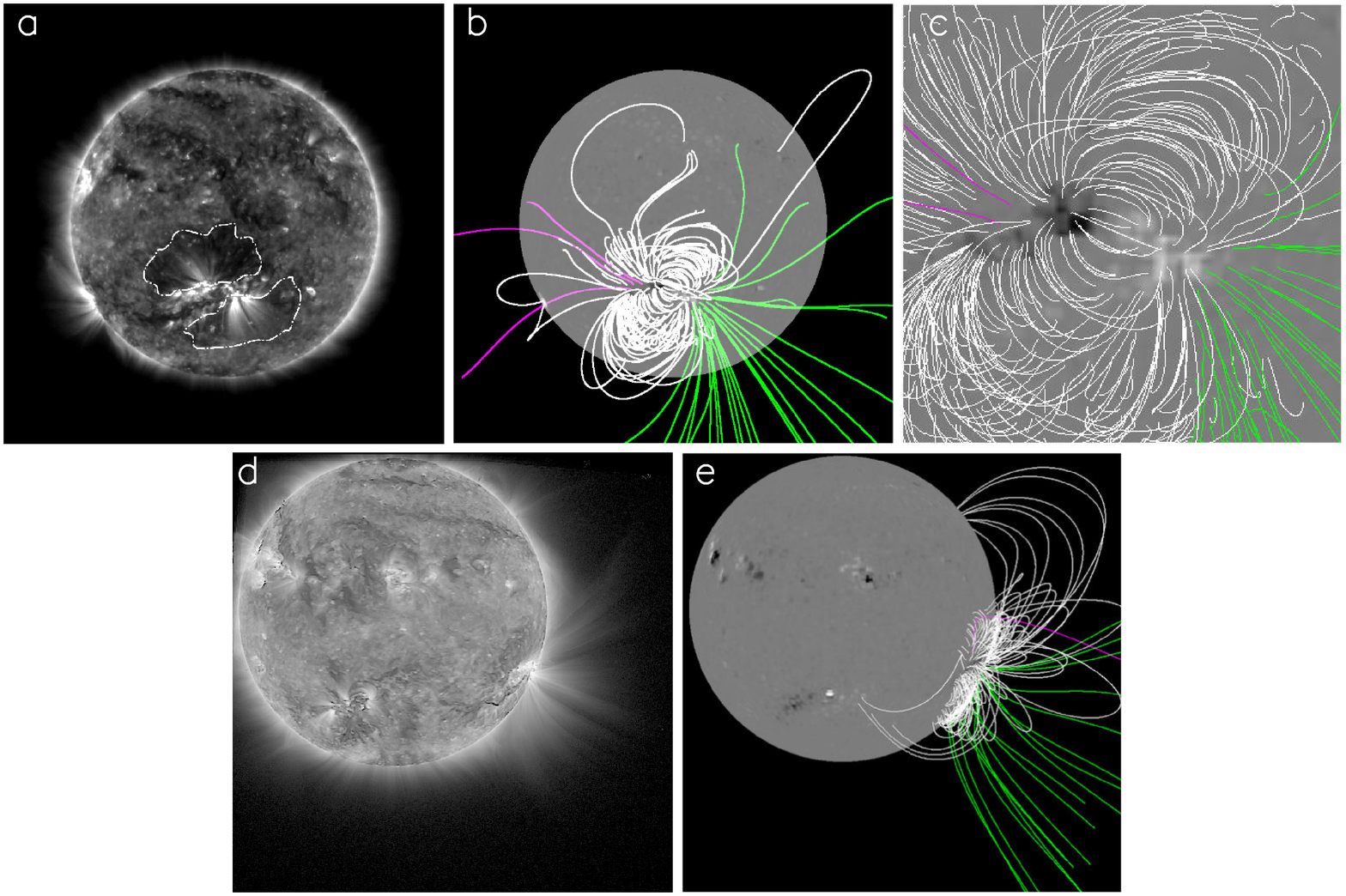}
\caption{The SWAP images of AR 11112 seen at the disk center on 14 October 2010 at 12:01:01~UT (a) and the extended corona seen above the AR at the western limb on 20 October 2010 at 01--04h UT (d). Dot-dashed lines designate the boundaries of the neighboring coronal holes. The PFSS extrapolations of the magnetic field were computed for 14 October (b and c) and 20 October (e) 2010, 12:04~UT. Open field lines of positive/negative polarities are marked by green/magenta colors.\label{fig01b}}
\end{figure}

We analyzed the SWAP images of the extended coronal structure observed at the western solar limb on 20 October 2010 between 01:06:46 and 03:55:26~UT, and on 21 October 2010 between 00:40:26 and 03:04:46~UT (see Fig. \ref{fig01b}). In both cases the cadence was 30~s. This structure originated above AR 11112, which was located near disk center around one week earlier (on 14 October 2010; Fig.~\ref{fig01b}a). The AR was positioned between two regions of reduced brightness, marked by dot-dashed lines. The Potential Field Source Surface (PFSS) extrapolation model \citep[we used Version 1 of the SSW PFSS package described in][]{DeRosa} shows that the magnetic configuration of these regions contains patches of open magnetic field lines of positive and negative polarities next to the AR (Figures \ref{fig01b}b and \ref{fig01b}c), which can be identified as small coronal holes in hotter lines with {\sl SDO}/AIA. Other parts of the outlined dark regions, aside from the patches of radially directed open magnetic field lines, may be identified as dark canopies filled with dense chromospheric material \citep{Wang11}. A large filament is seen to the southeast of the AR center indicating that the neutral magnetic field line crossed the meridian plane in the center of the AR at an angle of $\sim$~65$^{\circ}$. The fan rays expanding from both sides of the AR along open field lines suggest a magnetic configuration favorable to the presence of compact outflows \citep{Slemzin12}.

On 20 October 2010, when the AR was located at the western limb, it produced a rich structure of diverging coronal rays expanding to a distance of 2$R_{\odot}$, well aligned to the magnetic field lines computed from the PFSS solution (Fig.~\ref{fig01b}d and e). The northern part of the coronal structure at position angles $\lesssim 100^{\circ}$ (measured clockwise from the northern pole) corresponds to extended but closed loops. The central and southern parts of the structure (at larger position angles) correspond to open field lines.

\begin{figure}    
\epsscale{1.0}
\plotone{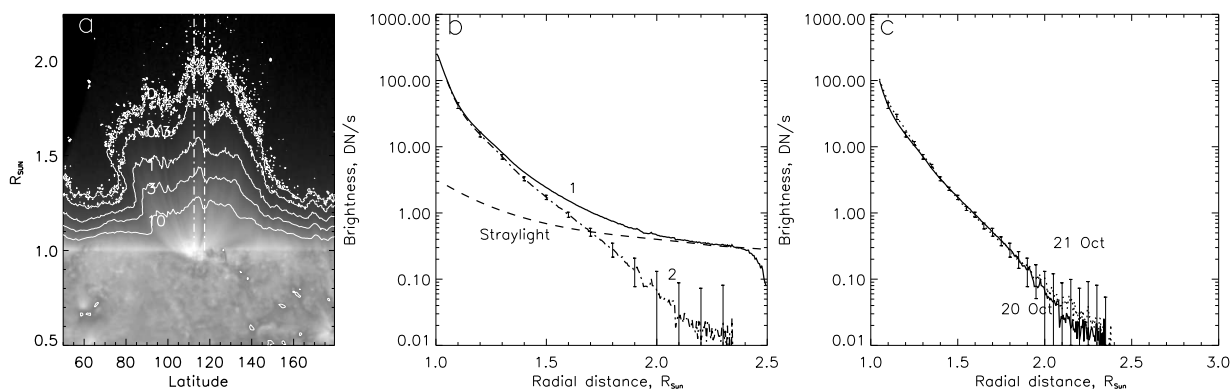}
\caption{(a) Polar diagram of the coronal structure shown in Fig.~\ref{fig01b}d. The contours map the levels of brightness from 10 to 0.1~DN. The dot-dashed lines mark out the boundaries of the analyzed coronal ray at the angle of 115$\pm$2.5$^{\circ}$; (b) the measured radial distributions of brightness in the ray before (curve 1) and after (curve 2) subtraction of the stray light (dashed line). The error bars include statistical errors of coronal brightness in the ray and in the stray light; (c) Superposition of brightness distributions in the ray derived from the data on 20 and 21 October 2010.\label{fig01c}}
\end{figure}

For our analysis, we used the level-1 SWAP FITS files, which were processed to correct for dark current, detector bias, flat-field variations, and bad pixels. The solar images were centered, rotated so that solar north pole was at the top position, re-scaled to a square-pixel format, and exposure time-normalized, so that individual pixel values were expressed in units of DN~s$^{-1}$ \citep{Seaton12}. For correct removal of stray light, we selected images (about 80 on each date) corresponding to the same LAR value at which the stray light distribution was measured (LAR=1). The selected images were co-added for each day and transformed to polar coordinates by linear interpolation. In the polar images, we selected a strip, covering position angles 115$\pm$2.5$^{\circ}$, that contains the longest and brightest radially directed ray in the center of the structure (Fig.~\ref{fig01c}a). Its radial brightness distribution was obtained by averaging over the strip width (Fig.~\ref{fig01c}b, curve 1), which was chosen to be sufficiently narrow not to cross the cavity above the filament, seen in the synoptic map of Fig.~\ref{fig05}(a) (see below).

The main obstacle to observing the corona at large heliocentric distances is the level of stray light in the telescope. Stray light in SWAP is produced by several sources, including small-angle scattering by irregularities of the mirror surface, diffraction by the entrance filter grid and intrinsic scattering of the back-reflected focused light by the rear filter, the first one likely being the main source. Due to the asymmetric optical design, stray light in SWAP solar images is also anisotropic \citep{Seaton12,Halain12}. Stray light is commonly described by the wings of the instrument's point spread function (PSF), which can be obtained from solar eclipse images using, for example, the blind deconvolution method \citep{Shearer12}. Determination of the PSF shape in the far wings is seriously complicated by photon and dark-current noise, which limit the effective distance over which deconvolution can be used, determined by a certain threshold of signal. The recent version of the SolarSoft package \citep{Freeland98} for SWAP contains an optional procedure to correct stray light by deconvolution with a predefined PSF. However, this procedure cannot remove stray light at distances above 1.5$R_{\odot}$, where the signal drops below 1~DN, so we have used an alternative method, which is free from this limitation.

\begin{figure}    
\epsscale{1.0}
\plotone{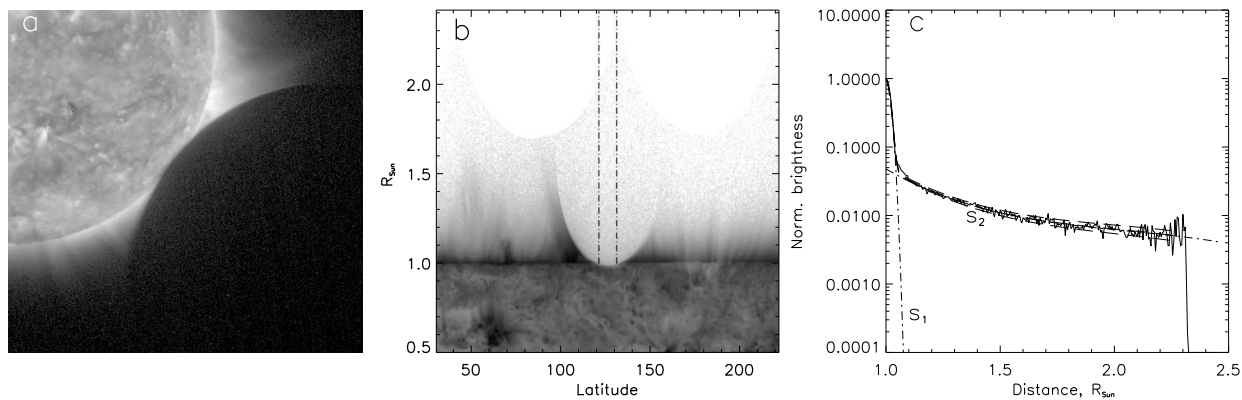}
\caption{Determination of the SWAP stray light distribution from the eclipse data obtained on 1 July 2010 at 07:55:38~UT: (a) - the original image; (b) - its polar transformation with the selected region at the angle of 125$\pm$5$^{\circ}$ marked by dot-dashed lines; (c) - a comparison of the measured stray light radial distribution with its analytical approximation consisting of two components: a Gaussian function S$_1$ and a cubic polynomial exponent S$_2$. Dashed lines mark the 1$\sigma$ error levels.\label{fig01a}}
\end{figure}

To determine stray light in the direction of the coronal ray, we analyzed the SWAP image obtained during the eclipse on 1 July 2011 at 07:55:38~UT, when the Moon obscured the corona up to the limb in the southwestern sector (Fig.~\ref{fig01a}a). After transformation into polar coordinates (Fig.~\ref{fig01a}b), we measured the residual brightness along the strip in the deepest part of the lunar shadow at the latitudinal angle of 125$\pm$5$^{\circ}$. This direction is close to the position of the coronal ray we analyze at angle 115$\pm$2.5$^{\circ}$. In principle, the signal in the obscured image contains, in addition to stray light, a contribution from the afterglow of the scintillator deposited on the CMOS detector from the previous, unobscured coronal image, but in our case this component is negligible due to the afterglow's very short decay time (1--3 milliseconds) with respect to the 30-s image cadence \citep{Seaton12}.

The normalized radial distribution of the signal averaged over the strip width is shown in Figure~\ref{fig01a}c. This distribution can be approximated by the sum of short-range (S$_1$) and long-range (S$_2$) components. The short-range component is not pure stray light. It represents the variation of the limb brightness outside the edge of the Moon superimposed on stray light inside the edge when the Moon shifts during the exposure time (10~s). This component can be approximated by Gaussian function with a half-width of about 10 pixels (32~arcsec), which constitutes $\approx$~0.03$R_{\odot}$. After subtraction of the short-range component, we evaluated the long-range component of the signal, which corresponds to stray light produced by scattering of the focused image by high-spatial frequency irregularities on the mirror's surface \citep{Harvey12} with a minor contribution due to the intrinsic back-reflected light \citep{Halain12}. This component is well modeled by a cubic polynomial exponent.

To obtain the radial dependence of stray light along the streamer, the long-range component was normalized to the signal at a distance of 2.5$R_{\odot}$, where we expect stray light to dominate. This procedure slightly overestimates stray light (to less than 10\%) at distances above 2$R_{\odot}$, but we did not use this region for analysis of coronal radiation due to large statistical errors. Curve 2 in Fig.~\ref{fig01c}b shows the resulting brightness distribution in the coronal ray after subtraction of stray light. The error bars correspond to the square root from the quadrature sum of the standard deviation of coronal brightness (across the strip) and the stray light uncertainty. Figure~\ref{fig01c}c presents the radial brightness distributions along the ray for 20 and 21 October 2010, which coincide within the measuring errors. As can also be seen from Figure~\ref{fig01c}b, the slope of the brightness curves grows below 1.2$R_{\odot}$, which may be the result of the superposition of many closed loops at the streamer base.

\begin{figure}    
\epsscale{1.0}
\plotone{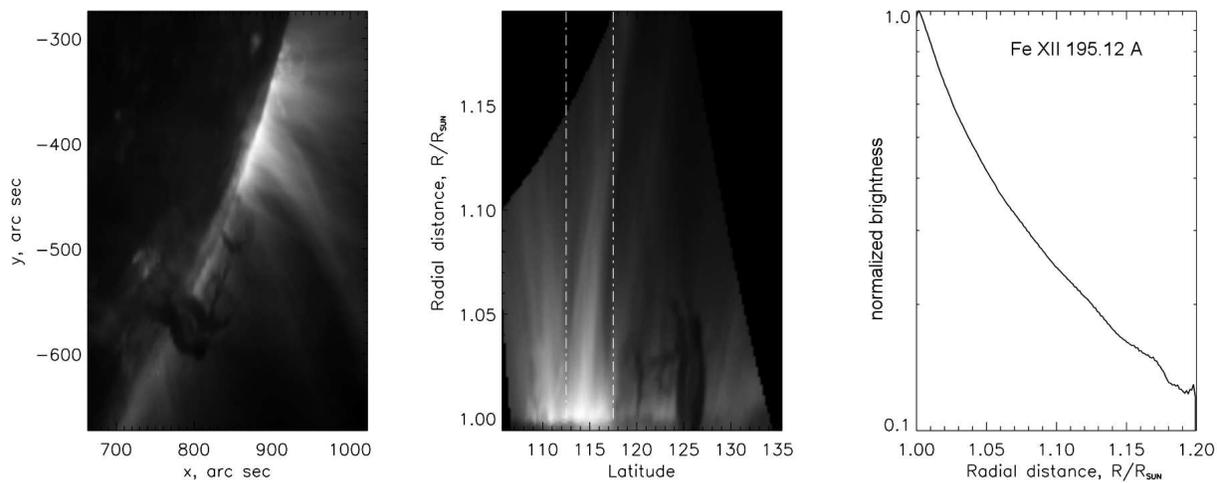}
\caption{The Hinode/EIS image of the active region AR 11112 in Fe~XII 195.12~\AA\ line observed on 21 October 2010 at 18:28:13~UT. Right is the original image, middle is its polar transformation with the analyzed coronal ray marked by the dot-dashed lines, left is the obtained radial intensity distribution in the coronal ray.\label{fig04}}
\end{figure}

In order to use plasma diagnostics to probe the base of the streamer, we analyzed {\sl Hinode}/EIS spectrometer data obtained during the specially coordinated EIS-SWAP campaign (HOP 165) on 21 October 2010 at 18:28:13~UT, when AR 11112 rotated to the western limb. After standard processing of the initial EIS FITS data, images in lines of Fe~{\sc ix} -- Fe~{\sc xv} were extracted and again transformed into polar coordinates. Figure~\ref{fig04} displays, as an example, the image of AR 11112 in the Fe~{\sc xii} 195.12~{\AA} line and its polar transformation. The streamer ray under study is bounded by dot-dashed lines as in Figure~\ref{fig01c}a. The radial brightness distributions in all spectral lines were obtained in a similar way as for the SWAP images. They were then corrected for stray light by subtracting a constant level as is recommended in the EIS data analysis documentation (EIS Software Note 12, distributed in {\sl SolarSoft}).

In order to interpret the derived distribution of brightness in the streamer ray, in the next two sections we will consider mechanisms of EUV emission in the corona and the physical conditions in the coronal plasma.

\section{Formation of the coronal EUV emission in Fe~{\sc ix} -- Fe~{\sc xi} lines}

The four spectral lines in SWAP's wavelength band are resonance lines, except the Fe~{\sc x} $\lambda$177.24 line. Two main excitation mechanisms of EUV Fe~{\sc ix} -- Fe~{\sc xi} lines in the solar corona are: (i) collisional excitation due to electron-ion collisions; and (ii) resonant scattering of photons coming from the lower corona. The local emission $E_{ki}$ [phot cm$^{-3}$ s$^{-1}$ sr$^{-1}$] of an EUV line emitted due to transition $i\to k$ can be then written in the form:

\begin{equation}
E_{ki}=\frac{1}{4\pi} \left( \varepsilon_{ki}^{\mathrm{(col)}} + \varepsilon_{ki}^{\mathrm{(scat)}}\right) \, , \label{tot_rate}
\end{equation}
where $\varepsilon_{ki}^{\mathrm{(col)}}$ and $\varepsilon_{ki}^{\mathrm{(scat)}}$ (in units of phot cm$^{-3}$ s$^{-1}$) are the volume emissivities of the collisional and resonant components of $E_{ki}$, respectively.

For the solar coronal plasma the collisional excitation component $\varepsilon_{ki}^{\mathrm{(col)}}$ with a good accuracy can be expressed as

\begin{equation}
\varepsilon_{ki}^{\mathrm{(col)}}=  N_k\, N_e\, {Br}_{ik}\, C_{ki}(T) \, , \label{col}
\end{equation}
where $N_k$ is the number density of ions in the ground state, $N_e$ the electron density number, $C_{ki}(T)$ the collisional excitation rate coefficient for the transition $k\to i$ calculated on the assumption of Maxwellian velocity distribution of plasma electrons at temperature $T$, ${Br}_{ik}$ the branching ratio coefficient equal to unity for resonance lines.

The emissivity of the resonant scattering component, $\varepsilon_{ki}^{\mathrm{(scat)}}$, is given by the expression \citep[see, e.g.,][]{Noci87}

\begin{equation}
\varepsilon_{ki}^{\mathrm{(scat)}}= N_k  B_{ki} \int_{\Omega} p(\theta) F(\delta\lambda) d\Omega  \, , \label{tot_excit}
\end{equation}
where $B_{ki} = (\pi e^2 /m c) f_{ki}$ is the Einstein coefficient for absorption ($f_{ki}$ the absorption oscillator strength for the transition $k\to i$), $p(\theta)$ is the angular coefficient taking into account the anisotropy of resonant scattering process ($\theta$ the angle between the direction ${\bf n'}$ of the radiation emitted from the solar disc and the direction ${\bf n}$ along the line of sight toward the observer), and $d\Omega$ is the element of the solid angle $\Omega$ subtended by the solar disc at a given ion's position in the solar atmosphere. The quantity $F(\delta\lambda)$ has the following form

\begin{equation}
F(\delta\lambda) = \int_0^{\infty} I_{\mathrm{ex}}(\lambda,{\bf n'}) \psi(\lambda - \lambda_0) d\lambda \, ,
 \label{convol}
\end{equation}
where $I_{\mathrm{ex}}(\lambda,{\bf n'})$ is the spectral intensity of incident radiation coming from the lower corona, $\psi(\lambda-\lambda_0)$ the normalized absorption profile ($\lambda_0$ the wavelength at the line center), and $\delta\lambda= (\lambda_0 / c) \mathbf{v}\cdot {\bf n'}$ the Doppler shift of the absorption profile with respect to that of the illuminating flux $I_{\mathrm{ex}}$ due to macroscopic movement of the coronal plasma with the radial expansion velocity $\mathbf{v}$. As was determined by \citet{Gimenez07}, the major portion of EUV emission in the spectral lines under consideration is generated in the lower corona at the heights of about 7~Mm.

For determining the resonant scattering rate (\ref{tot_excit}), we use the following assumptions: (i) the mean value for the angular factor $p(\theta)$ is taken equal to $p=1/4\pi$, a good approximation for this factor \citep[see][]{Noci87}; (ii) the flux of incident radiation $I_{\mathrm{ex}}$ does not depend on the direction ${\bf n'}$. In these assumptions the emissivity of the resonant component can be written as follows \citep[see, e.g.,][]{Gabriel71}

\begin{equation}
\varepsilon_{ki}^{\mathrm{(scat)}}= N_k 4\pi \frac{\pi e^2}{m c} f_{ki} \frac{\lambda_0^2}{c} F_{ki} W(r) D(T_1,T_2,\mathbf{v}) \, ,
 \label{scat_rate}
\end{equation}
where $F_{ki}$ [phot cm$^{-2}$ s$^{-1}$ sr$^{-1}$] is the mean intensity of emission of the underlying corona in the spectral line of wavelength $\lambda_{ki}$, $W(r)=(1- ( 1-1/r^2)^{1/2})/2$ is the geometric dilution factor obtained on the assumption of the uniform emission of the lower corona above the disc and accounting for the weakening of the solar surface radiation at heliocentric distance $r$. The quantity $D(T_1,T_2,\mathbf{v})$ called the Doppler dimming factor is proportional to the probability for the incident photon emitted by the lower corona to be scattered by an ion. It is the convolution product of the two profiles (see Eq. (\ref{convol})): the profile of the illuminating radiation $\phi(\lambda)$ with the temperature $T_1$, and that of absorbing coronal plasma $\psi(\lambda-\lambda_0)$ with the temperature $T_2$ and bulk velocity $\mathbf{v}$.

\section{Physical conditions in the corona and modeling of EUV emission}\label{sect4}

To compute the EUV emission of coronal plasma in the selected spectral lines, we need to know the physical conditions in the plasma: electron density and temperature; ion and elemental abundances; the bulk outflow velocity of the plasma; and the kinetic temperature of coronal ions. These parameters combined allow us to describe line intensities and profiles as a function of  radius or along the line of sight, and are usually derived from observational spectroscopic data, using various plasma-diagnostic techniques.

A brief review of density and temperature measurements in a number of studies of the corona in polar and equatorial regions was presented by \citet{Andretta12}. It is worth noting some of those studies, which we compare below with the results of our work. \citet{Saito77} derived equatorial and polar electron-density models of quiet corona from measurements of polarized brightness in white light. \citet{Gibson99} modeled electron densities within a streamer region using both white-light and EUV observations. \citet{Antonucci06} obtained the electron-density distribution of the slow coronal wind flowing along the streamer boundaries. \citet{Thernisien06} performed a 3D reconstruction of the density structure of a streamer using observational data from {\sl SOHO}/LASCO. The models of the electron temperature distributions in streamers along radial distance were obtained by \citet{Gibson99} from white-light density profiles and by \citet{Vasquez03} from the {\sl SOHO}/UVCS data.

Recently, a detailed study of physical conditions in a coronal prominence cavity above a filament and surrounding streamer observed on 9 August 2007 was published in a series of papers: \citet{Gibson10}, \citet{Schmit11}, and \citet{Kucera12}. In the paper by \citet{Gibson10}, the morphological parameters of the model, in which the cavity is considered as a tunnel-like low density tube with elliptical cross section in a coronal streamer, were determined from the {\sl STEREO}/EUVI data. Data from {\sl Hinode}/EIS and the MLSO/Mk4 coronagraph were then used by \citet{Schmit11} to forward-model the EUV and white-light data, in order to derive electron densities as a function of altitude in both the cavity and the streamer. Lastly, \citet{Kucera12} used previous results and EIS observations in a set of iron lines to forward-model the temperature profile in the cavity and streamer. The range of altitudes in those works was limited, with values $R = 1.05-1.2 R_{\odot}$.

The atomic data needed to calculate emissivities in the spectral lines produced by collisional excitation were taken from the CHIANTI atomic database version 7.1 \citep{Dere97,Landi13}. To estimate the radiative component produced by resonant scattering (see Eq. (\ref{scat_rate})), we used the line intensities $F_{ki}$ determined from the SWAP and EIT images. At first, the SWAP image taken on 14 October 2010, -- when the base of the streamer was located at disk center -- was segmented by brightness into quiet regions, active regions and coronal holes; the mean signals in DN/pix for each region were then determined. The theoretical intensities of the individual lines in each region were computed with the CHIANTI package using the corresponding canonical DEM functions provided in the database, with the Fe abundance taken from \citet{Feldman92} and the ionization equilibrium taken from \citet{Mazzotta98}. The supposed SWAP signal (in units of DN) in a given line was determined from its theoretical intensity using the SWAP spectral calibration given by \citet{Raftery13}. The calculated signals in different lines were summed for each type of region, and the total calculated signal was compared with the measured mean value. The ratio of the measured-to-calculated signal was used as the normalization coefficient for any given region. Finally, the intensities in each line, averaged over the whole solar disk, were determined by the expression:

\begin{equation}
F_{ki} = \sum_{m=1}^{3} I_{ki}(m)\, n_m \, A_m \, ,
 \label{dimming}
\end{equation}
where $I_{ki}(m)$ is the calculated intensity for a given line in the region $m$ (quiet Sun, active region, or coronal hole), $n_m$ is a normalization coefficient, and $A_m$ is the relative area of the selected region $m$ on the Sun. The results of the calculation are shown in Table \ref{T-simple1}. For verification, the same procedure was made for the EIT 171~\AA\ image taken on the same day. The differences in the flux values are less than 50\%, well within the typical accuracy of the EIT spectral calibration \citep{Dere00}. In calculations of the resonantly scattered component of the coronal emission we assume that the EUV brightnesses are uniformly distributed over the disk area. This assumption is justified by the fact that, at solar minimum, the distribution of brightness over the solar disk in the SWAP 1~MK spectral band is rather homogeneous. The reason for this is that local structures like active regions and coronal holes make a weaker contribution to the total flux than the sum of the uniformly distributed flux from quiet regions: active regions produce less than 30\% and coronal holes less than 10\% of the total flux. Naturally, images of the disk in hotter lines are less homogeneous. Our estimation showed that under this assumption, the application of the dilution formula for the 171~\AA\ spectral band at solar minimum allows us to estimate the EUV flux illuminating the corona at heliocentric distances $R> 2 R_{\odot}$ with an accuracy of better than 100\%, which is comparable with the SWAP calibration errors.

Using the approach described above, we can make preliminary estimates of the relation between radiative and collisional excitation rates in the corona. Taking the typical Saito density model for the quiet equatorial corona \citep[see][]{Saito77} and using Eqs. (\ref{col}) and (\ref{scat_rate}), the radiative component for the strongest Fe~{\sc ix} 171.08~\AA\ line becomes comparable with the collisional component at radial distances $R > 2.5 R_{\odot}$. For other lines in the SWAP spectral band, the radiative component is less than for this particular line. As will be shown below, in the streamer we study, this threshold distance is even higher.

\begin{figure}    
\epsscale{0.6}
\plotone{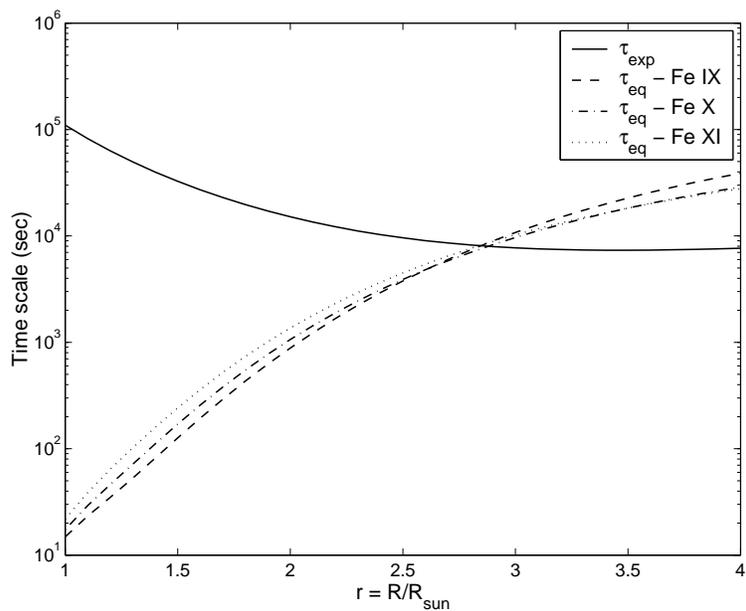}
\caption{Characteristic times as a function of heliocentric distance according to plasma expansion velocity model of \citet{Withbroe82}.\label{fig_timescale}}
\end{figure}

In addition, there is the important question as to whether thermodynamic equilibrium conditions are satisfied in the corona in the presence of plasma bulk flows. Because of the relatively low densities in the upper regions of the solar atmosphere, various types of particle may have different temperatures, and ionization equilibrium may also not correspond to the local electron temperature. In order to find out at which distances these effects become important, we compared the characteristic times for coronal expansion and the establishment of thermal and ionization equilibrium. To evaluate the expansion time of the slow wind plasma, $\tau_{\mathrm{exp}}(r)=[ (v/n_e)\, dn_e/dr ]^{-1}$, we used the radial flow-velocity model $v(r)$ from \citet{Withbroe82} and the density distribution model $n_e(r)$ of the equatorial background corona from \citet{Saito77}. The slow wind velocity model of \citet{Withbroe82} gives velocities of about 10--20~km~s$^{-1}$ in the distance range 1.5--2.2$R_{\odot}$. The ionization and recombination rates used in the estimation of the ionization equilibrium timescales, $\tau_{\mathrm{eq}}$, for Fe~{\sc ix} -- Fe~{\sc xi} ions were derived from the CHIANTI database using temperature distribution models by \citet{Gibson99} and \citet{Vasquez03}. The results of this analysis are presented in Figure \ref{fig_timescale}. We found that the timescales for Fe~{\sc ix} -- Fe~{\sc xi} ions ionization equilibrium $\tau_{\mathrm{eq}}$ are one order of magnitude less than the plasma expansion time $\tau_{\mathrm{exp}}$ in the corona up to a distance of $R \approx 2.2 R_{\odot}$, which suggests that thermal equilibrium exists across our whole region of interest. It is also worth noting that there exist studies \citep[e.g.,][]{Antonucci05,Antonucci06}, where slow plasma outflow velocities reach values up to 100~km~s$^{-1}$. Nevertheless, for the spatial region of our interest, the condition $\tau_{\mathrm{exp}} > \tau_{\mathrm{eq}}$ holds. We note that \citet{Antonucci05,Antonucci06} analyzed streamer boundaries and regions above the streamer cusp, while we consider the central bright streamer part, where outflow velocities are distinctly lower.

\section{Calculation of the streamer emission and comparison with measurements}

The distributions of plasma parameters along the streamer ray can be determined from the forward-modeling of coronal brightness. The relations in Sections 3 and 4 allow the calculation of coronal EUV emission from unit volume as a function of the electron density and temperature. To obtain the total emission along the line of sight we need to know the longitudinal density and temperature profiles of the ray. These spatial characteristics, in principle, can be determined from variation of coronal brightness at different heights using solar rotation, i.e. from rotational tomographic projections.

The method of three dimensional (3D) reconstruction of coronal structures from rotational tomographic projections was developed by \citet{Frazin05} and for the first time applied to the white-light LASCO C2 images \citep{Frazin02}. \citet{Vasquez11} and \citet{Nuevo12} used this method in the EUV range for the 3D reconstruction of the local differential emission measure in the nearest to the limb region of the corona ($R$=1~--~1.25~$R_{\odot}$) based on the {\sl STEREO}/EUVI and {\sl SDO}/AIA images. The principal limitations of current implementations of this method are: a limited angular resolution (2$^{\circ}$ in both latitude and longitude); and high sensitivity to temporal variations of brightness. Thereby, the applicability of this method to retrieve the shape of coronal structures near active regions needs special consideration.

\begin{figure}    
\epsscale{1.0}
\plotone{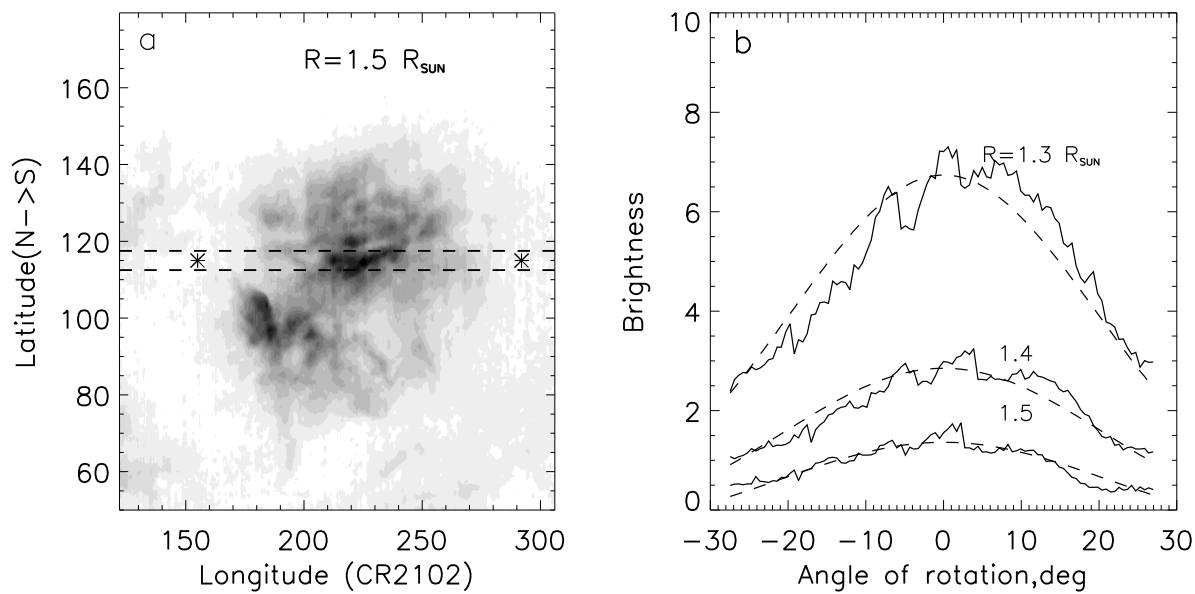}
\caption{(a) Part of the SWAP synoptic map. The horizontal dashed lines mark the boundaries of the streamer ray. The asterisks correspond to the background corona on the same latitude as the ray. (b) Solid lines show the brightness variations in the streamer with the rotational angle determined from the synoptic map at different heights, dashed lines are the corresponding best-fit curves.\label{fig05}}
\end{figure}

Figure \ref{fig05}(a) shows a part of the SWAP synoptic map of the corona for the Carrington rotation 2102 containing the region under investigation. The coronal ray we analyze is bounded by dashed lines. The light curves in Figure \ref{fig05}(b) display the variations in line-of-sight brightness of the ray at three heights: 1.3, 1.4, and 1.5~$R_{\odot}$. Simple modeling of a ray as a static slab gives a smooth sinusoidal dependence of line-of-sight brightness on the rotational angle shown by the dashed curves. In fact, the real light curve (marked by the solid lines) shows significant modulation of brightness correlated across all three heights sampled. The modulation peaks have widths of 2--3 degrees, and their amplitudes exceed the statistical errors by a factor of 2--3. Probably, this modulation can be produced by temporary flows of plasma existing for several hours superimposed on the main body of the streamer. According to \citet{Vasquez11}, when the measured intensity displays such temporal variability, the method of one-dimensional rotational tomography cannot give reliable results. Better results can be derived from multi-point tomography with sufficient temporal resolution, but this is beyond the scope of this paper.

In absence of information about the streamer's profile, to model its EUV emission we used a different approach based on estimation of the effective extent of its density distribution function along the line of sight, based on photometric analysis. We define the effective depth (or width) of the streamer ray along the line of sight (LOS), $L_{\mathrm{eff}}$, in a given EUV spectral line by the expression $L_{\mathrm{eff}}(\lambda)=B_{\mathrm{LOS}}(\lambda)/E_{\lambda}(T,N_e)$, where $B_{\mathrm{LOS}}(\lambda)$ is the measured brightness (expressed in units of phot cm$^{-2}$ s$^{-1}$ sr$^{-1}$) in the spectral line, and $E_{\lambda}(T,N_e)$ is the local emission (see Eq.~(\ref{tot_rate})) calculated from the independently determined values of density $N_e$ and temperature $T$ of the plasma at the same point in the corona.

\begin{figure}    
\epsscale{0.9}
\plotone{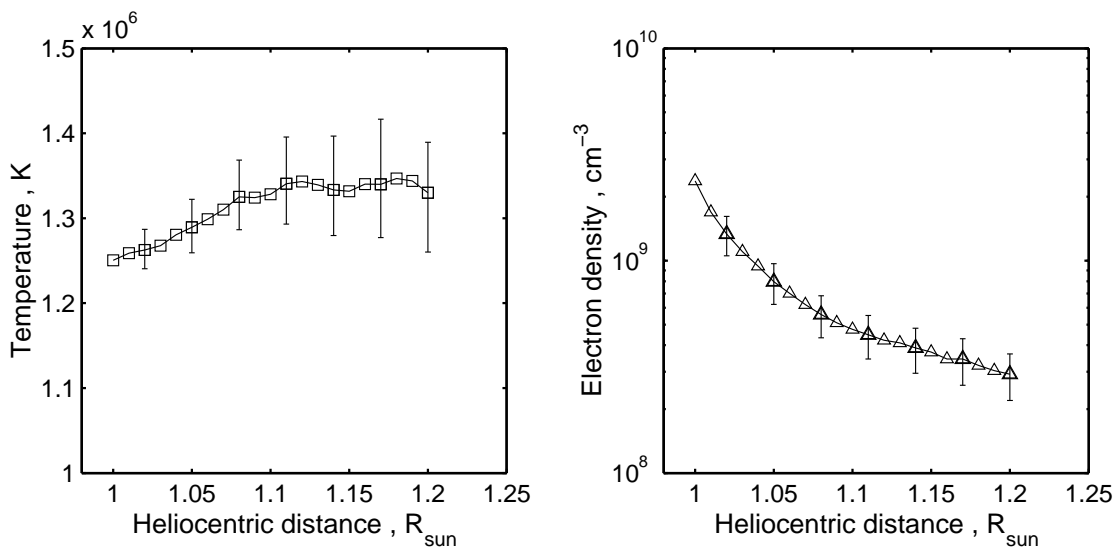}
\caption{The temperature (left panel) and density (right panel) radial distributions derived from the EIS data.\label{fig05a}}
\end{figure}

At the base of the streamer ray at $R\le 1.2 R_{\odot}$, the plasma parameters (electron density and temperature) were derived from the {\sl Hinode}/EIS data (Section \ref{sect2}) using a line-ratio diagnostic technique. For determining the contribution functions of EIS EUV lines (see Eq. (\ref{eff_length}) below) we used the CHIANTI version 7.1 database (see Section \ref{sect4} for details). The plasma temperature was evaluated from the line-intensity ratio Fe~{\sc x} $\lambda$184.54~\AA /(Fe~{\sc xi} $\lambda$188.23~\AA +Fe~{\sc xi} $\lambda$188.30~\AA ). It was found out that in the range $R=1-1.2R_{\odot}$ the temperature $T_e$ increased smoothly with radius from $\approx$~1.25~MK to 1.35~MK. Electron densities, derived from the line ratio (Fe~{\sc xii} $\lambda$186.85~\AA + Fe~{\sc xii} $\lambda$186.88~\AA )/Fe~{\sc xii} $\lambda$195.12~\AA , decreased from $\approx$2$\cdot$10$^9$~cm$^{-3}$ to $\approx$3$\cdot$10$^8$~cm$^{-3}$. The results of temperature and density diagnostics are shown in Figure \ref{fig05a}. The errors shown in these graphs include the statistical errors on the initial data and an assumed 20\% accuracy for the CHIANTI calculations for intensity of each line.
We also estimated the effect of the reduction in EIS's sensitivity over time \citep{DelZanna13} on the ratios of the EIS line intensities under consideration, and found that this effect does not noticeably change the resulting density and temperature values when compared with the measurement errors on the observed data.

The plasma parameters obtained through diagnostics were used to determine the effective width $L_{\mathrm{eff}}$ at heliocentric distance 1.2$R_{\odot}$, taking into account the uncertainties associated with line intensities, atomic data, and electron density and temperature errors. The values of $L_{\mathrm{eff}}$ (in units of $R_{\odot}$) at the point 1.2$R_{\odot}$ found from the EIS data were: in the Fe~{\sc x} $\lambda$184.54 line -- 0.138$\pm$0.053, the Fe~{\sc xi} $\lambda$188.23 -- 0.133$\pm$0.046, the Fe~{\sc xii} $\lambda$186.85 -- 0.140$\pm$0.051, and the Fe~{\sc xii} $\lambda$195.12 -- 0.145$\pm$0.050.

In order to forward-model the EUV emission, we used the following assumptions about the streamer and surrounding coronal structure:

\smallskip

\noindent 1. We suppose that the streamer is embedded in the background corona, so that the electron density in each point of the line of sight can be described as a sum of two components:

\begin{equation}
N_e(x,y,z) = N_e^{(\mathrm{str})}(x,y,z) + N_e^{(\mathrm{bgr})}(x,y,z) \, ,
 \label{dens_structure}
\end{equation}
where $N_e^{(\mathrm{str})}$ is the component of coronal density corresponding to the streamer, and $N_e^{(\mathrm{bgr})}$ the density distribution of the spherically symmetric background corona. For the coordinates ($x,y,z$) the following directions are chosen: (i) $z$ is the radial distance from the Sun's center, projected to the plane of the sky and directed along the coronal ray under study, (ii) $x$ is the axis along the line of sight, and (iii) $y$ is the coordinate perpendicular to the $z$ axis in the plane of the image.

\smallskip

\noindent 2. The streamer density distribution along the line of sight can be described by the slab model similar to that used by \citet{Guhathakurta96}, \citet{Vibert97}, and \citet{Thernisien06} to model white-light streamers:

\begin{equation}
N_e^{(\mathrm{str})}(x,y,z) = N_e(z)\, F_{\|}(x,z)\, F_{\bot}(y) \, ,
 \label{distr_dens}
\end{equation}
where $N_e(z)$ gives the radial part of the electron density distribution along the $z$ axis, $F_{\|}(x,z)$ is the function characterizing the shape of the streamer along the line of sight, $F_{\bot}(y)$ the function representing variations of density in the plane of the solar image. In Eq. (\ref{distr_dens}) the $F_{\bot}(y)$ part is assumed to be averaged within the limits of the ray strip for a given coordinate $z$, so that the density distribution (\ref{dens_structure}) is considered hereinafter as a function of variables $x$ and $z$, i.e. $N_e(x,z)$.

\smallskip

\noindent 3. The density distribution in the streamer along the line of sight, $F_{\|}(x,z)$, is described by a Gaussian function

\begin{equation}
F_{\|}(x,z) =  \exp\left[ - \left(
\frac{x}{s_e(z)}\right)^2 \right]\, ,\label{distr_dens2}
\end{equation}
where the streamer width $s_e(z)$ characterizes the transverse dimension of the streamer and is associated with the dimensional parameter $L_{\mathrm{eff}}$ (see below). For calculations of the integrated brightness in the streamer above 1.2$R_{\odot}$ we consider the streamer as a radially diverging slab with a constant angular width $\alpha$ relative to the solar center.

\smallskip

\noindent 4. The plasma temperature in the streamer is assumed to be constant along the line of sight. The background corona is supposed to be isothermal with the temperature equal to the mean temperature in the streamer.

\smallskip

According to Eq.~(\ref{dens_structure}), the brightness of the corona along the line of sight can be similarly presented as the sum of two components

\begin{equation}
B_{\mathrm{LOS}}(\lambda ,z) =  B_{\mathrm{str}}(\lambda ,z) + B_{\mathrm{bgr}}(\lambda ,z)  \, ,\label{tot_bright}
\end{equation}
where $B_{\mathrm{str}}$ and $B_{\mathrm{bgr}}$ correspond to brightness of the streamer and background corona, respectively. This means that for a correct analysis of EUV emission of the streamer we need to take into account the contribution of the background corona to the total brightness. The latter was extracted from the SWAP images taken on 14 and 27 October 2010 (see asterisks in Figure \ref{fig05}a), with the condition that the line-of-sight brightness at the latitude range of the streamer ray (namely 115$\pm$2.5$^\circ$) was minimal over the period studied.

When collisional excitation dominates and there is constant temperature along the $x$ direction, the intensity of an optically thin EUV line of wavelength $\lambda$ for the streamer's part can be expressed as

\begin{equation}
I_{\mathrm{str}}(\lambda ,z)=\int G\left(\lambda,T(z)\right) \, \left(N_e^{\mathrm{(str)}}(x,z)\right)^2\, dx  \, ,
\label{eff_length}
\end{equation}
where $G\left( \lambda,T(z)\right)$ [phot cm$^3$ s$^{-1}$ sr$^{-1}$] is the temperature-dependent contribution function (for the resonance Fe ion lines under consideration, $G$ is not sensitive to the variations in electron density experienced in the corona), for which we adopt the following form:

\begin{equation}
G\left(\lambda,T(z)\right) = \frac{1}{4\pi} \frac{N_j(\mathrm{Fe}^{+m})}{N(\mathrm{Fe}^{+m})}
\frac{N(\mathrm{Fe}^{+m})}{N(\mathrm{Fe})} \frac{N(\mathrm{Fe})}{N(\mathrm{H})}
\frac{N(\mathrm{H})}{N_e} \frac{A_{ji}}{N_e} \, ,
\label{G_func}
\end{equation}
where \, $A_{ji}$\, is the Einstein spontaneous emission coefficient for the transition $j\to i$, $N_j(\mathrm{Fe}^{+m})$/$N(\mathrm{Fe}^{+m})$ is the fraction of ions $\mathrm{Fe}^{+m}$ at level $j$, $N(\mathrm{Fe}^{+m})/N(\mathrm{Fe})$ is the fraction of Fe ions with charge ${+m}$, $N(\mathrm{Fe})/N(\mathrm{H})$ is the abundance of iron relative to hydrogen, and $N(\mathrm{H})/N_e$ is the number density ratio of hydrogen nuclei to electrons. The contribution functions for the SWAP lines were calculated using the CHIANTI version 7.1 \citep{Dere97,Landi13}. We used the coronal Fe abundance of \citet{Feldman92}, the ionization equilibrium of \citet{Mazzotta98}, and a hydrogen-to-electron density ratio of 0.83.

In order to obtain the total EUV brightness, $B_{\mathrm{str}}$, in the SWAP spectral band, the partial contribution functions for individual lines should be convolved with the SWAP response coefficients $Q_{\lambda}$ (given in Table \ref{T-simple1}) and summed to give the SWAP temperature response function:

\begin{equation}
\widetilde{G} (T(z)) = \sum_{\lambda} G\left(\lambda,T(z)\right)\cdot  Q_{\lambda}   \, .\label{response}
\end{equation}
The modeled SWAP brightness of the streamer (in units of DN s$^{-1}$) is then given by

\begin{eqnarray}
B_{\mathrm{str}}(z) = \int \widetilde{G}\left(T(z)\right) \, \left(N_e^{\mathrm{(str)}}(x,z)\right)^2\, dx \nonumber \\
= \widetilde{G}\left( T(z)\right)\cdot EM(z)  \, ,
\label{eff_length2}
\end{eqnarray}
where $EM(z) = \langle (N_e^{\mathrm{(str)}})^2(z)\rangle_x \cdot L_{\mathrm{eff}}(z)$ is the emission measure of plasma along the line of sight. As according to  Eq. (\ref{eff_length2}) $B_{\mathrm{str}}\propto N_e^2$, the distribution of brightness along the line of sight has a width $s_b$ satisfying the relation $s_e = \sqrt{2} s_b$ (see Eq. (\ref{distr_dens2})). If we define a relationship between $s_b$ and $L_{\mathrm{eff}}$ as $L_{\mathrm{eff}}=2 s_b$, the Gaussian width $s_e$ can be then obtained from the expression $s_e = L_{\mathrm{eff}}/\sqrt{2}$. Under the assumption of a divergent slab model for the streamer with a constant vertex angle $\alpha$, we have $s_e(z) = \sqrt{2}\, z\, \tan (\alpha /2)$.

\begin{figure}    
\epsscale{1.1}
\plottwo{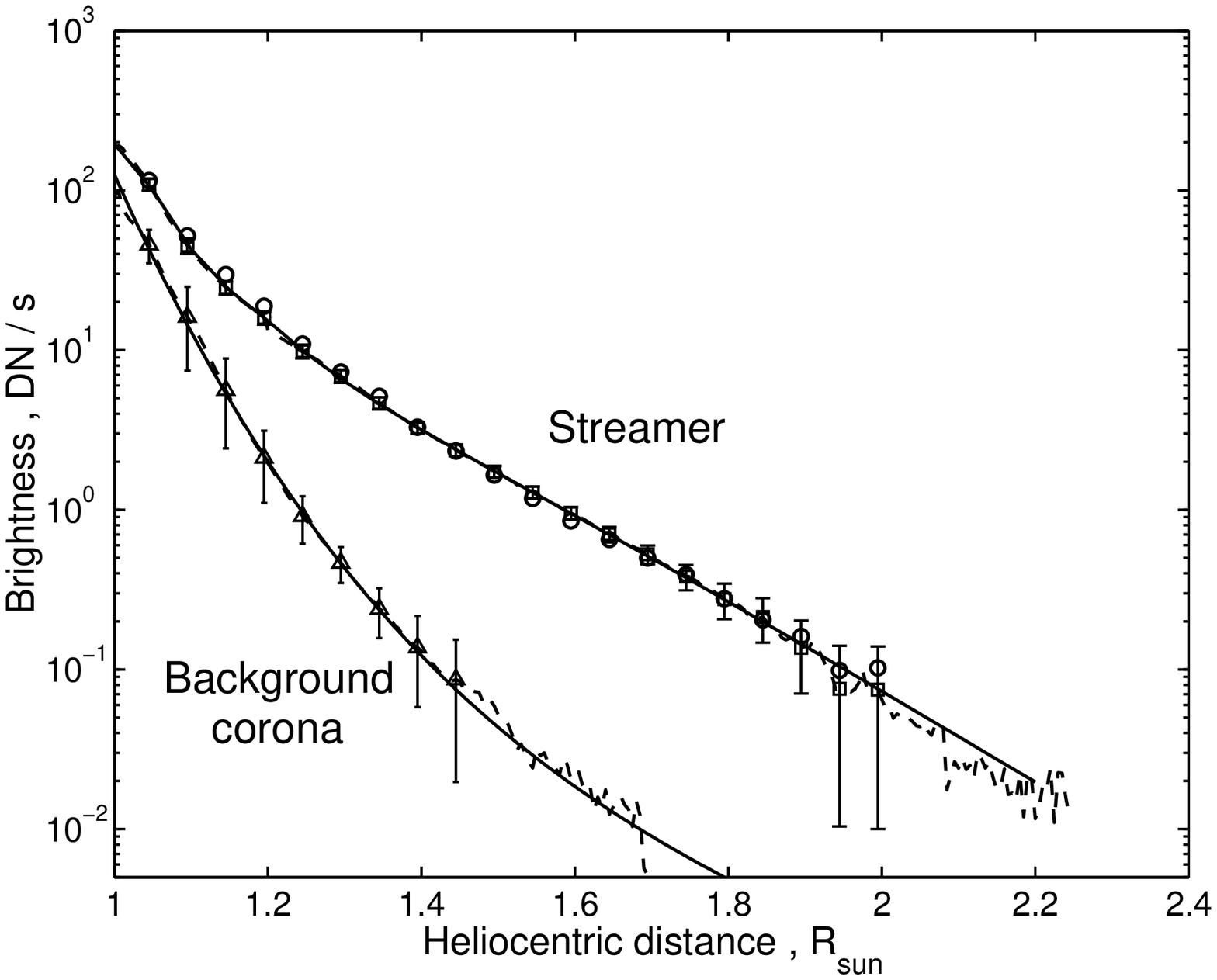}{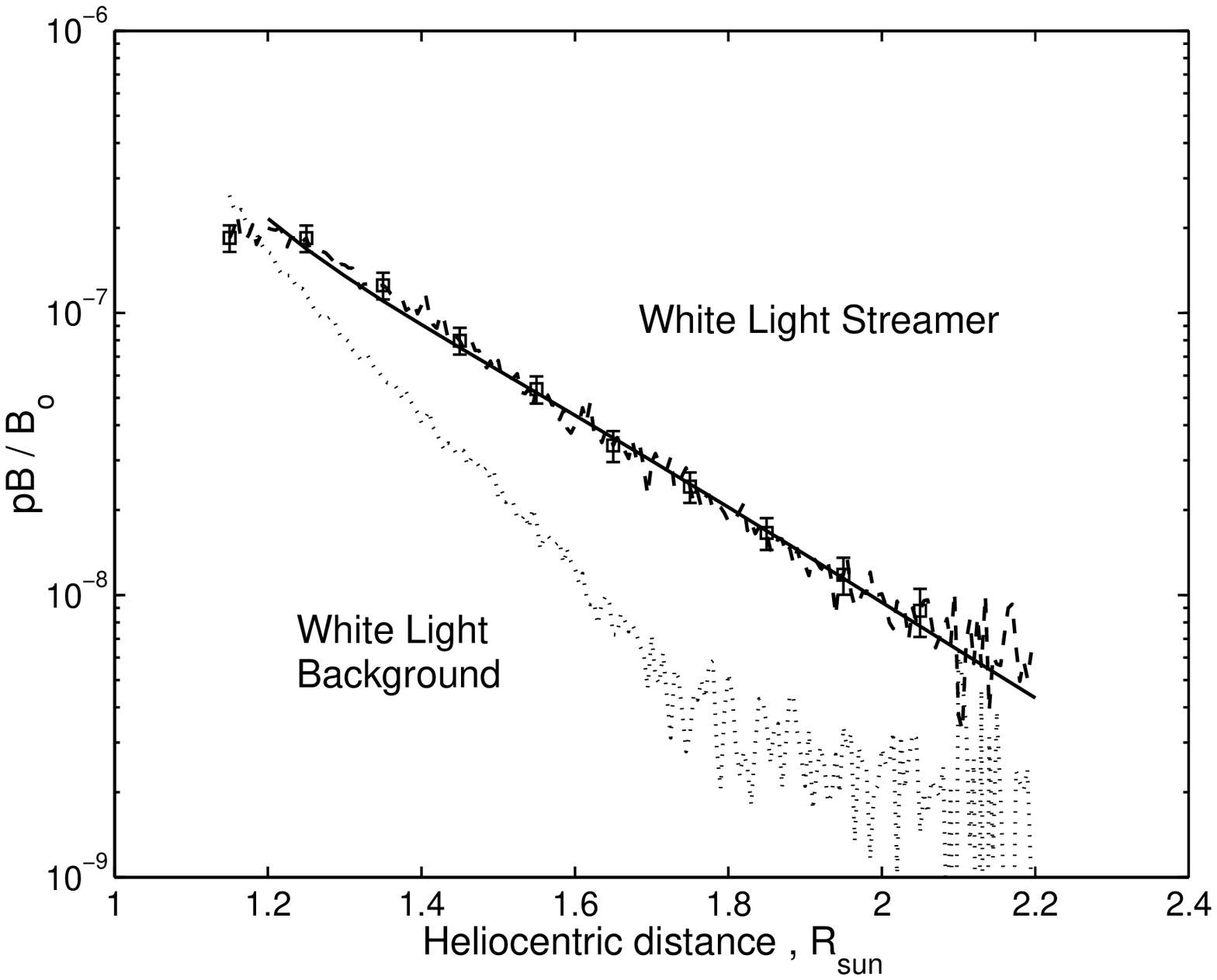}
\caption{Left: EUV brightness of the streamer and background corona determined from the SWAP observations on 20 and 21 October 2010 in comparison with the modeled ones using the found density and temperature distributions. Dashed curves are observational data for 20/10/2010 (square error bars) and background corona (triangle error bars); circles correspond to the observational data for 21/10/2010; solid curves present the modeled brightness distributions. Right: Comparison of the model with the MLSO/Mk4 coronagraph data on 20 October 2010 ($B_\mathrm{o}$ is the mean brightness of the solar disk). Dashed line with square error bars is the MLSO/Mk4 data on 20/10/2010; solid curve is the modeled polarization brightness $pB$; dotted curve corresponds to the background corona on 14 October 2010.\label{fig06}}
\end{figure}

According to Eq. (\ref{eff_length2}), for modeling the observed coronal EUV brightness, we need to know two quantities: emission measure (or density) and temperature. In addition to the SWAP data, for determining the electron density in the streamer, we used the independent data for the white-light coronal brightness, measured by the MLSO Mk4 coronagraph \citep{Elmore03} on 20 October 2010. For analysis we made use of the data on polarization brightness, $pB$, which corresponds to Thomson scattering of photospheric white light from free electrons in the corona at distances from $\approx$1.15 to 2.2$R_{\odot}$. The value of $pB_{\mathrm{str}}$ for the streamer part is then given by the integral along the line of sight:

\begin{equation}
pB_{\mathrm{str}}(z) = \int_{\mathrm{LOS}}  C(r,\theta) \, N_e^{(\mathrm{str})}(r) \, dx \,\, ,
 \label{pB}
\end{equation}
where $C(r,\theta)$ is the Thomson scattering function ($r$ is the heliocentric distance to scattering point, $\theta$ the angle between the line of sight and $r$-direction), and $N_e^{(\mathrm{str})}(r)$ is the density distribution given by Eqs. (\ref{distr_dens}) and (\ref{distr_dens2}). The brightness of the background corona in visible light was determined from the Mk4 data taken on 14 October 2010. Figure \ref{fig06} displays the resulting brightness distributions in the streamer ray and in the background corona in EUV (left panel) and in visible light (right panel).

To calculate the radial density distribution $N_e(z)$ at the streamer core from the Mk4 data, we forward-modeled the polarized brightness in white light according to Eq. (\ref{pB}), assuming the streamer is a diverging slab with a Gaussian distribution of density along the $x$-direction. The width of the slab at the reference point 1.2$R_{\odot}$ was found by fitting the density modeled from the Mk4 data in accord with the value obtained from the EIS line ratio using a width of the Gaussian function (\ref{distr_dens2}) (which is equivalent to $L_{\mathrm{eff}}$) as a fitting parameter. We consider the EIS density value reliable for two reasons. First, it was derived from the ratio of the Fe~{\sc xii} line intensities which are insensitive to temperature. Second, the brightness of the background at the reference point was small in comparison with that of the streamer (less than 10\%), so its presence has very little influence on the observed line ratio. Thus, the density and temperature values at 1.2$R_{\odot}$ obtained from the EIS line ratios can be reasonably attributed to the streamer.

\begin{figure}    
\epsscale{1.1}
\plottwo{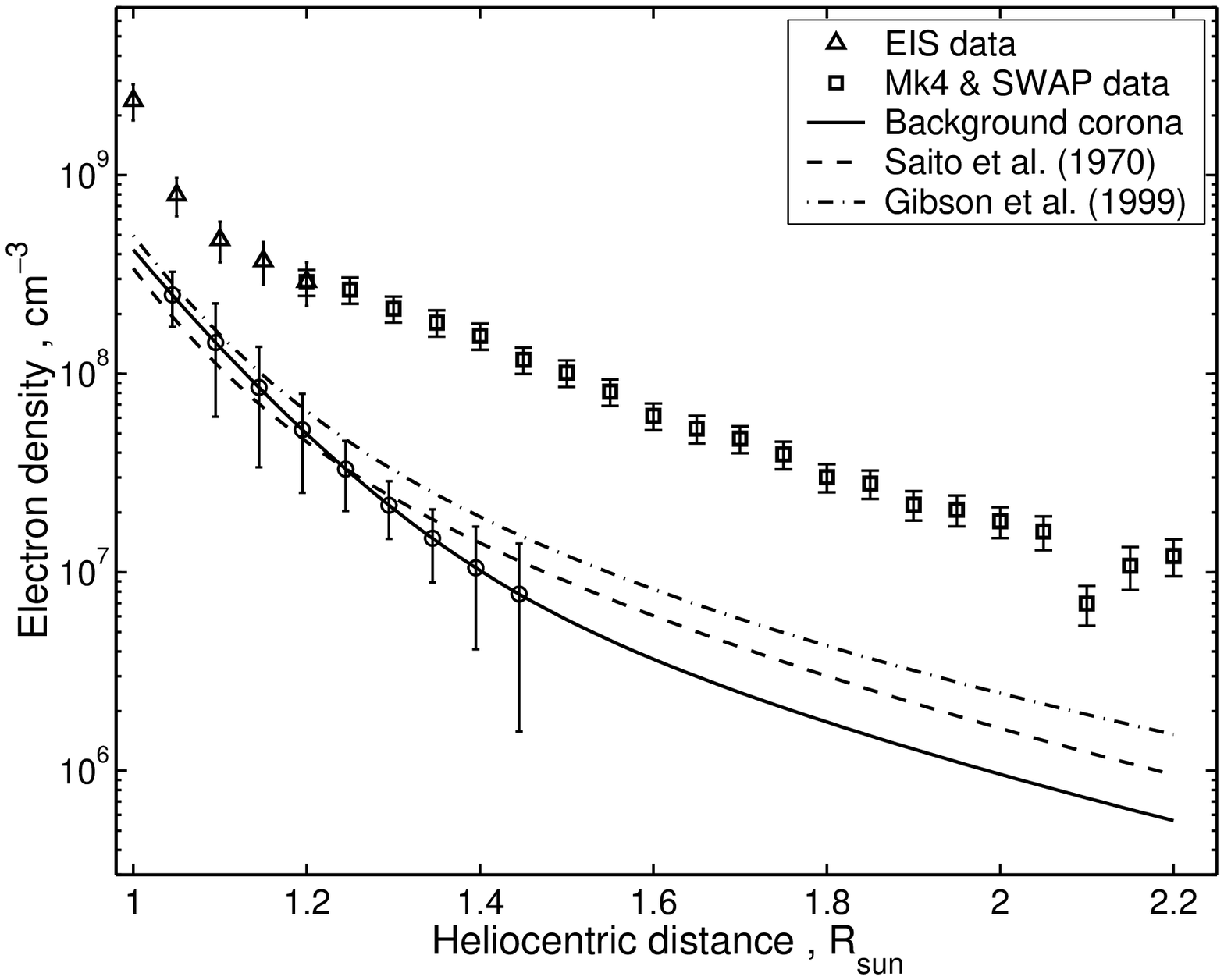}{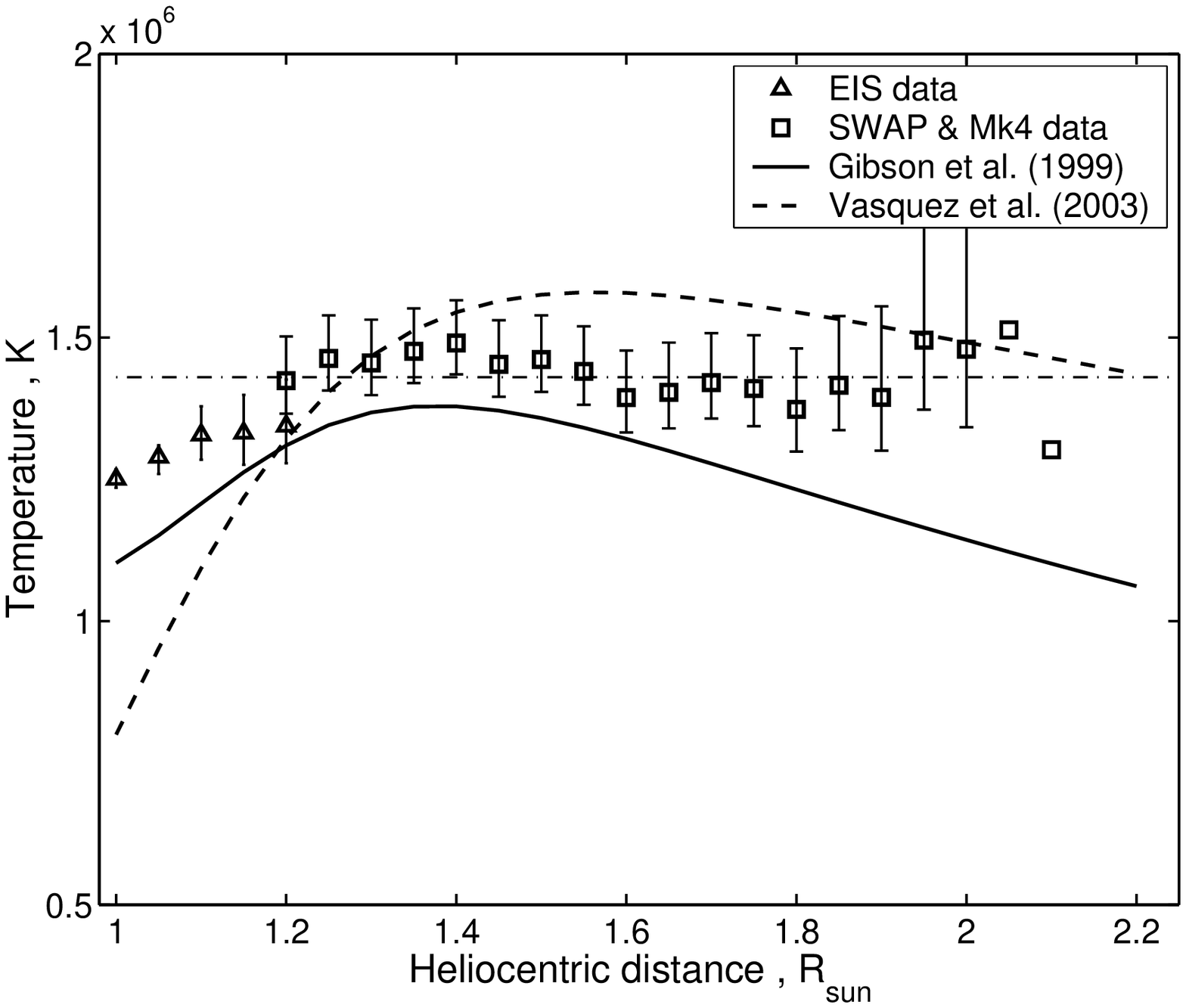}
\caption{Left: Radial distributions of mean density in the streamer for 20/10/2010 (derived from EIS and Mk4 \& SWAP data) and background corona (solid line denoted by error bars below 1.6$R_\odot$ and extrapolated to distances above 1.6$R_\odot$), and their comparison with other model density data. Right: Radial distributions of mean temperature in the streamer for 20/10/2010 and its comparison with the data of other authors. Dash-dotted horizontal line corresponds to the mean temperature $T$=~1.43~MK.\label{fig07}}
\end{figure}

The best-fit value $L_{\mathrm{eff}}=0.139$ in the reference point (which corresponds to the angular width of the slab $\alpha\approx 6.6^{\circ}$) obtained from the modeling of the polarization brightness turned out to be in good agreement with the values derived from the EIS line intensities, lying in the range 0.133--0.145. The resulting density distribution in the streamer shown in Figure \ref{fig07} (left panel) was approximated in a logarithmic form by the polynomial:

\begin{equation}
\log_{10} N_e(z) = \sum_{i=0}^3 \frac{a_i}{z^{i}} \, , \label{dens1}
\end{equation}
The coefficients of this distribution are presented in Table \ref{T-parameters}.

Using the derived density distribution $N_e(z)$, the radial dependence of plasma temperature in the streamer above 1.2$R_{\odot}$ was determined by forward-modeling the EUV emission in the SWAP spectral band to best fit the measuring data. According to preliminary estimates of the excitation rates (see Chapter 4), in our model we took into account only EUV emission due to collisional excitation. The modeled EUV brightness in the streamer was determined from Eq. (\ref{eff_length2}) as a function of temperature, making use of the emission measure calculated from the Mk4 data. The radial dependence of temperature obtained from a comparison of the modeled EUV brightness with measurements is shown in Figure \ref{fig07} (right panel) combined with the EIS line ratio values at $R < 1.2 R_{\odot}$. At the point $R = 1.2 R_{\odot}$ the EIS temperature may be slightly underestimated, probably, due to a larger influence of straylight at the edge of the EIS field of view than is prescribed in the EIS documentation. Finally, the plasma temperature in the streamer's stalk varies within the range 1.3--1.5~MK with deviations from the mean value of about 1.43~MK, no more than the determination errors. So, within the accuracy of 0.08~MK the streamer plasma above 1.2$R_{\odot}$ can be regarded as isothermal, i.e. $T=1.43\pm 0.08$~MK. The modeled EUV brightness distributions for the streamer and coronal background (left panel), as well as $pB$ radial dependence in the streamer (right panel), calculated with the found values of density and temperature are shown in Figure \ref{fig06} by the solid curves.

We also evaluated the electron density distribution in the background corona using the SWAP measurements (see the left plot in Figure \ref{fig06}) under the assumption that it is spherically symmetric and isothermal with a mean temperature in the streamer of $T$=~1.43~MK. Electron densities were determined from the SWAP EUV data using a technique similar to that developed by \citet{Hulst50} for a spherically symmetric corona, but adopting a quadratic dependence of brightness on density. The background density distribution is shown in Figure \ref{fig07} by the solid curve at the left panel. It was obtained by extrapolation above 1.6$R_\odot$ because of the lack of observational EUV data for the background (see Figure \ref{fig06}). This distribution was approximated with formula similar to (\ref{dens1}), where the variable $z$ has to be replaced with the heliocentric distance $r$. Parameters of the approximation are also presented in Table \ref{T-parameters}. The density in the background corona was about one order of magnitude or even more less than that in the streamer. For comparison, Figure \ref{fig07} shows the density and temperature distributions obtained by \citet{Saito70}, \citet{Gibson99}, and \citet{Vasquez03}. Our results for the electron density distribution in the background corona are consistent with the results of these authors at distances below 1.5$R_\odot$ and a little diverge above this distance.

We have analyzed whether the derived density distribution in the streamer ray conforms to a hydrostatic density model

\begin{equation}
 N_e(r) = N_e(r_1) \exp \left[ - \frac{(r-r_1)}{h_0 r_1 r}  \right] \, , \label{dens2}
\end{equation}
where $N_e(r_1)$ is the electron density at the reference point $r_1$, $h_0 = kT/(\overline{m} g_{\odot}\zeta r_1)$ is the density scale height expressed in units of the solar radius $R_{\odot}$, $\overline{m}$ -- the mean atomic mass of particles in the corona, $g_{\odot}$ -- the solar surface gravitational acceleration, $\zeta$ -- the ratio of the thermal gas pressure to the total effective pressure \citep{Orrall90}. The value $T_{\mathrm{sh}}=T/\zeta$ is defined as the ``scale height temperature''. According to \citet{Guhathakurta92}, $h_0 = 0.0729 T_{\mathrm{sh}}$~[MK].

\begin{figure}    
\epsscale{0.6}
\plotone{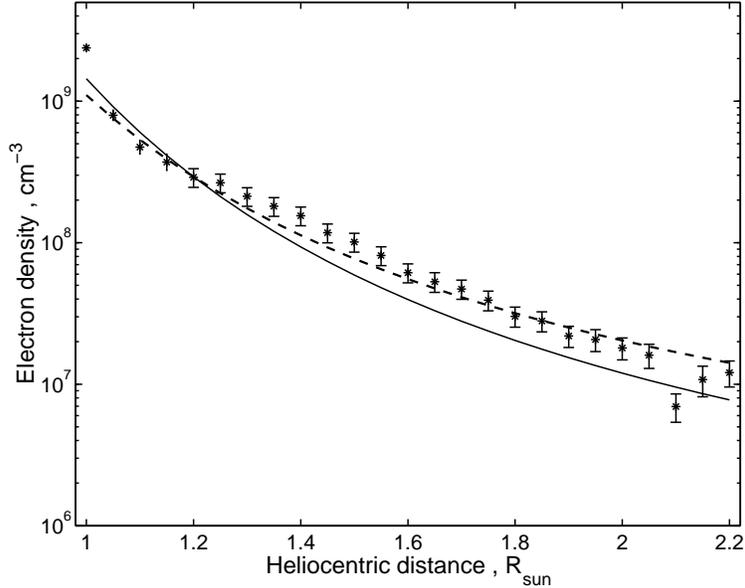}
\caption{Comparison of the present modeled density distribution (asterisks with error bars) with hydrostatic density model at $T_{\mathrm{sh}}=1.43$~MK (solid line) and $T_{\mathrm{sh}}=1.72$~MK (dashed line).\label{fig_hydro}}
\end{figure}

In Figure \ref{fig_hydro} we show the hydrostatic density distribution calculated with the scale-height temperature equal to the mean temperature in the streamer $T_{\mathrm{sh}}=1.43$~MK and normalized to the measured density at the reference point $r_1=1.2 R_{\odot}$. The measured densities are higher by a factor of 1.7 with respect to hydrostatic model values, the difference being four times more than the mean error in density ($\sim$17\%). The scale-height temperature determined as the best fit to the measured data is equal to 1.72$\pm$0.08~MK, which is sufficiently higher than the temperature obtained from our photometric model. It suggests that the plasma in the streamer contains a non-thermal component raising the total effective gas pressure some $\sim$70\% above the hydrostatic equilibrium value. These results are similar to those obtained by \citet{Warren02} who analyzed a quiet coronal streamer at lower distances -- from 1.05 to 1.35$R_{\odot}$ -- and found that over the observed heights the plasma in the streamer was nearly isothermal with $T=1.45$~MK being ``overdense'' relative to the hydrostatic equilibrium by a factor of 1.52. Probably, the excess of density in the streamer is caused by an outward bulk plasma flow, which may contribute to the solar wind.

\begin{figure}    
\epsscale{0.9}
\plotone{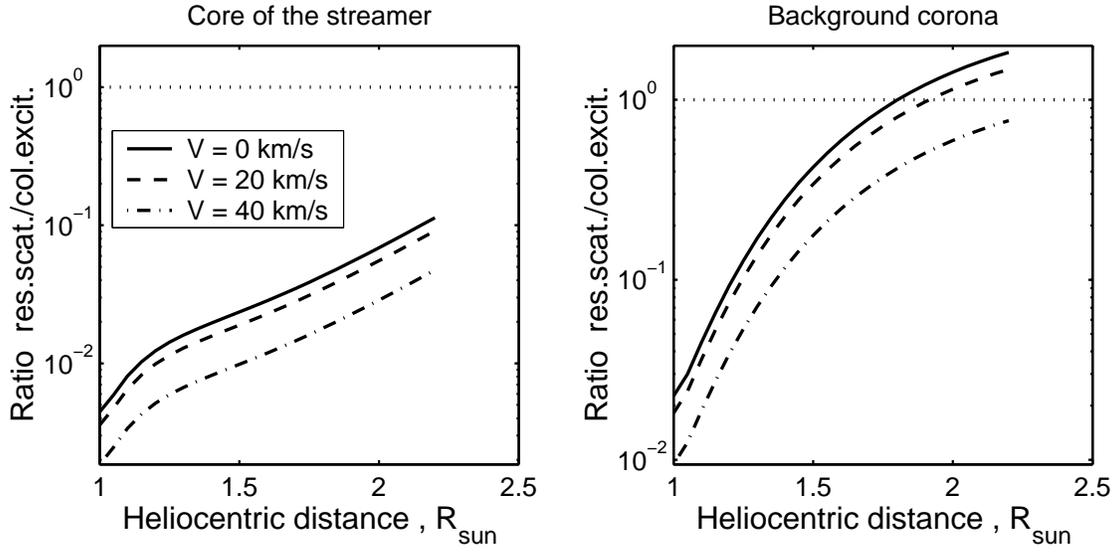}
\caption{Ratio of the resonant scattering rate to the collisional excitation rate for Fe~IX 171.08~\AA\ line at $T=$~1.4~MK and electron densities from the present work (see Eq. (\ref{dens1}) and Table \ref{T-parameters}) as a function of the heliocentric radius and the radial flow velocity $v$.\label{fig_ratio}}
\end{figure}

Using the derived density values for the streamer and backgound corona, we estimated the relative contribution of the resonant excitation mechanism to line formation in the SWAP EUV spectral band. Figure \ref{fig_ratio} shows the ratio of the resonant (radiatively excited) component of coronal EUV emission to the collisional one in our derived density model for the streamer ray and background corona, in an elementary volume of plasma, as a function of heliocentric radius $R$. Results are shown for both static and radially moving plasmas. As is seen from Figure \ref{fig_ratio}, in the streamer the resonant component contributes less than 10\% of the emission below $R\sim 2.5R_{\odot}$. This inference is in agreement with the assumption of \citet{Habbal11}. However, in the background corona the resonant component becomes comparable with the collisional one at heliocentric distances $\sim 2R_\odot$. In both cases, the contribution of the resonant component decreases with increasing bulk velocity of the coronal plasma through the Doppler dimming effect.

\section{Summary and conclusions}

This paper presents the results of a photometric study of {\bf a} streamer observed on 20--21 October 2010 with the SWAP EUV telescope in the 174~\AA\ spectral band, together with the {\sl Hinode}/EIS spectrometer (obtained in the HOP 165 session) and the Mauna Loa Mk4 white-light coronagraph. The application of special observational modes -- off-pointing, summation of several dozen high-cadence images and stray-light correction -- have allowed us to obtain the radial distribution of EUV brightness in the longest coronal ray to a distance of more than $2R_\odot$ with acceptable signal-to-noise ratio.

In order to understand the origin of the EUV emission observed by SWAP, we considered two main mechanisms of excitation of the Fe~{\sc ix} -- Fe~{\sc xi} ion lines that contribute to the flux in the SWAP spectral band: collisional excitation of ions by electrons and resonant scattering of the monochromatic radiation generated in the underlying corona. The incident fluxes in these spectral lines were determined from the SWAP and EIT images using segmentation of the structure types on the solar disk and the calculation of their component fluxes, using the CHIANTI package. Assuming a commonly used radial density distribution in the quiet equatorial corona \citep{Saito70}, an estimate of the relative contributions of collisional and resonant excitation to the EUV emission has shown that for the strongest Fe~{\sc ix} 171.08~\AA\ line, collisional excitation dominates up to radial distances in excess of 2.5$R_\odot$. For this range of distances, the characteristic timescale for reaching ionization equilibrium between the Fe~{\sc ix} -- Fe~{\sc xi} ions is shorter than the plasma expansion timescale \citep{Joselyn79,Withbroe82}, indicating that thermal equilibrium is reached in the streamer plasma.

The radial distributions of plasma density and temperature in both the streamer and the surrounding corona have been determined by forward-modeling their white-light and EUV emission, to best fit their measured values. The selected streamer ray was modeled as a slab with constant angular divergence from the solar center, a Gaussian density distribution and a constant temperature along the line-of-sight. In the first step, the plasma temperature and density distributions in the ray from the limb to the distance 1.2$R_\odot$ were determined from the EIS EUV data using Fe~{\sc x}/Fe~{\sc xi} and Fe~{\sc xii} ion line ratios, respectively. The effective width of the streamer along the line of sight, in these spectral lines, has been estimated as a ratio of the measured line brightnesses to their local emission, which were calculated using CHIANTI and the inferred plasma density and temperature. The electron density distribution along the coronal ray above the reference point, 1.2$R_\odot$, was then determined by forward-modeling the K-corona emission and fitting this to the polarization brightness measured by the Mk4 coronagraph. The parameters to be fitted were taken as the effective width of the streamer and density at the reference point derived from the EIS data. The validity of our model is confirmed by the fact that the effective width, which is derived independently from the white-light data (0.139$R_\odot$, or $\approx$~6.6$^{\circ}$ relative to the center of the Sun) is in good agreement with the values obtained from the EUV data (0.133--0.145$R_\odot$).

In the following step, we performed a forward-modeling of the streamer's EUV emission in the SWAP spectral band. A fitting of the modeled EUV brightness to the SWAP data gave the radial dependence of the product of emission measure (along the line of sight) with the temperature-dependent contribution function in the given Fe ion lines. Using the density values already determined from Mk4 data, we obtained the temperature distribution along the streamer. As a result, the forward-modeling of white-light and EUV brightnesses to match the measured data from all three instruments allowed us to obtain a self-consistent solution that yielded the most probable radial distributions of density and temperature in the streamer ray and surrounding background corona.

It was found that the plasma temperature in the streamer ray varied from 1.25~MK at the limb to 1.43~MK at 1.2$R_\odot$, being almost constant up to a distance of 2$R_\odot$. The density in the streamer ray varied from $\approx$~2$\cdot$10$^9$cm$^{-3}$ at the limb to $\approx$1--2$\cdot$10$^7$cm$^{-3}$ at 2$R_{\odot}$, and in the background corona it varied from $\approx$2--3$\cdot$10$^8$ to 2--3$\cdot$10$^6$ at the distance $\approx$1.7$R_{\odot}$. It was also found that the plasma density in the streamer decreases with distance sufficiently slower than it follows from the hydrostatic model with scale-height temperature equal to the independently determined value of 1.43$\pm$0.08~MK. The best-fit scale-height temperature to the obtained density distribution is found to be noticeably higher, at 1.72$\pm$0.08~MK. This result could mean that the plasma in the streamer contains a non-thermal component of motion, perhaps associated with an outward plasma flow. The existence of such time-varying plasma flows is consistent with the modulation of brightness in the SWAP images of the streamer during solar rotation.

Our photometric analysis has confirmed that enhanced brightness of ray-like coronal structures in Fe~{\sc ix} -- Fe~{\sc xi} EUV lines up to distances of $\sim 2 R_{\odot}$ can be explained by collisionally excited emission. Taking into account the plasma parameters determined, it was found that below this height, the resonant component constitutes less than 10\% of the total flux, whereas in the background corona at $\sim 2R_{\odot}$ it may in fact become comparable with contribution due to collisional excitation. However, the contribution of resonant scattering may be lower if the plasma moves outward with a velocity $\gtrsim$40~km~s$^{-1}$, due to the Doppler dimming effect.

In conclusion, our results show that the EUV emission of a streamer at distances from the limb up to $\approx 2R_\odot$ can be correctly modeled by assuming collisional excitation in a plasma in thermal equilibrium. We hope that studies of the EUV coronal emission at larger distances will be continued by forthcoming solar missions such as Solar Orbiter \citep{Muller13} and ASPIICS \citep{Lamy10}.

\acknowledgments
We would like to acknowledge Dr. D.~Berghmans, Dr. D.~Seaton and the ROB team for providing us with SWAP data and useful discussions. We are grateful to Prof. L.K.~Harra with her colleagues for organizing the coordinated Hinode/EIS and SWAP observations (HOP 165 session). The authors are thankful to Prof. L.~Golub for interest to our work and valuable advices. The research leading to these results has received funding from the European Commission's Seventh Framework Programme (FP7/2007-2013) under the grant agreement eHeroes (project No 284461, www.eheroes.eu), the PROBA2 Guest Investigator programme grant and partial financial support from Russian Foundation for Basic Research (grant 11-02-01079). SWAP is a project of the Centre Spatial de Liege and the Royal Observatory of Belgium funded by the Belgian Federal Science Policy Office (BELSPO). Hinode is a Japanese mission developed and launched by ISAS/JAXA, with NAOJ as domestic partner and NASA and STFC (UK) as international partners. It is operated by these agencies in co-operation with ESA and NSC (Norway). The Mk4 data were used by courtesy of the Mauna Loa Solar Observatory, operated by the High Altitude Observatory, as part of the National Center for Atmospheric Research (NCAR). NCAR is supported by the National Science Foundation. CHIANTI is a collaborative project involving George Mason University, the University of Michigan (USA) and the University of Cambridge (UK).




\clearpage



\clearpage

\begin{table}
\begin{center}
\caption{List of the SWAP spectral lines, mean solar line fluxes $F_{ki}$ (comparison of the SWAP and EIT data), and SWAP response $Q_{\lambda}$.\label{T-simple1}}
\begin{tabular}{lcccccc}     
\tableline\tableline
Ion & $\lambda$ , \AA\ &   $\log_{10} T_{\mathrm{max}}$ & $f_{ki}$ &  Solar flux, & phot cm$^{-2}$ s$^{-1}$ sr$^{-1}$ & SWAP response $Q_{\lambda}$,\\
    &          &          & &      SWAP        & EIT            & DN/ph cm$^2$ str pix \\
\tableline
Fe {\sc ix} & 171.08 &  5.9 &  2.946 & 1.1$\cdot$10$^{13}$ & 1.2$\cdot$10$^{13}$   & 1.63$\cdot$10$^{-12}$   \\
Fe {\sc x} & 174.53 &  6.0 &   1.252 & 4.3$\cdot$10$^{12}$ & 5.8$\cdot$10$^{12}$   & 1.97$\cdot$10$^{-12}$  \\
Fe {\sc x}& 177.24 &  6.0 &    0.708 & 4.5$\cdot$10$^{12}$ & 3.2$\cdot$10$^{13}$   & 1.38$\cdot$10$^{-12}$    \\
Fe {\sc xi} & 180.41 &  6.1 &  0.957 & 3.2$\cdot$10$^{12}$ &4.9$\cdot$10$^{12}$   & 3.01$\cdot$10$^{-13}$   \\
\tableline
\end{tabular}
\end{center}
\tablecomments{$T_{\mathrm{max}}$ is the temperature of the maximum abundance for each ion.}
\end{table}

\begin{table}
\begin{center}
\caption{Fitted parameters for the density distributions in the streamer
and background corona.\label{T-parameters}}
\begin{tabular}{lcccc}     
\tableline\tableline
Model & $a_0$ & $a_1$ &  $a_2$ & $a_3$  \\
\tableline
 Streamer  & 1.5996e+00 & 1.8606e+01 & -1.8230e+01 & 7.0165e+00 \\
 Background  & 4.4986e+00 & 1.0993e+00 & 4.3538e+00 & -1.3057e+00   \\
\tableline
\end{tabular}
\end{center}
\end{table}




\end{document}